# Quantifying coherence with principal diagonal elements of density matrix

Manis Hazra and Debabrata Goswami[†]
*Indian Institute of Technology Kanpur, Kanpur-208016, India*
[†]*Corresponding Author*: dgoswami@iitk.ac.in

Being the key resource in quantum physics, the proper quantification of *coherence* is of utmost importance. Amid complex-looking functionals in quantifying coherence, we set forth a simple and easy-to-evaluate approach: Principal diagonal difference of coherence ($C_{PDD}$), which we prove to be non-negative, self-normalized, and monotonic (under any incoherent operation). To validate this theory, we thought of a fictitious two-qubit system (both interacting and non-interacting) and, through the laser pulse-system interaction (semi-classical approach), compare the coherence evolution of $C_{PDD}$ with the relative entropy of coherence ($C_{r.e}$) and $l_1$-norm of coherence ($C_{l_1}$), in a pure-state regime. The numerical results show that the response of $C_{PDD}$ is better than the other two quantifiers. To the best of our knowledge, this letter is the first to show that a set of density-matrix diagonal elements carries complete information on the coherence (or superposition) of any pure quantum state.

*Introduction.—* Quantum coherence, one of the fundamental ingredients in quantum mechanics, determines the non-classicality of a physical system. However, a structured framework is a prerequisite for a better understanding of this important feature in a quantitative manner. In recent times, there has been a significant thrust in the setting up of Quantum resource theoretic models (QRT-models) in quantum coherence [1–4] or quantum superposition [5,6] as motivated by the resource theoretic model of quantum entanglement [7,8]. These developments have resulted in rigorous improvement in understanding the various research areas, like quantum thermodynamics [9,10], quantum metrology [11,12], quantum computation and communication [6,13,14], quantum biology [15,16], etc.

These resource-theoretic models are developed mainly by addressing three aspects: (a) characterization, (b) quantification, and (c) manipulation of states under constraints. To measure the coherence (superposition) of a quantum state, one needs a function (or functional) formed by the density-matrix elements of the state, called the coherence *quantifier*. Quantifiers satisfying the four necessary and sufficient criteria [1], namely, *positivity* (*faithfulness*), *monotonicity*, *strong-monotonicity,* and *convexity*, are highly sought after in the study of coherence or superposition (Coherence and superposition measures are very similar, except that while coherence demands an orthonormal basis, superposition only needs linear independence [6]. Thus, the superposition measure is a more generalized form. Since our theory is equally valid for coherence and superposition hereafter, we use the term *coherence* to mean both, if not mentioned apriori). Many quantifiers have been put forward thus far, like relative entropy of coherence [1], $l_1$-norm of coherence [1], trace norm of coherence [17,18], the geometric measure of coherence [19], skew information of coherence [20], the robustness of coherence [21], etc. Among these, the *relative entropy of coherence* ($C_{r.e}$) and $l_1$-*norm of coherence* ($C_{l_1}$) are well-verified candidates and attract attention from the scientific communities.

The *relative entropy of coherence* ($C_{r.e}$), predominantly an influential consequence of the *relative entropy of entanglement* [5,8], is simply stated as

$$C_{r.e}(\rho) = S(\rho_{diag}) - S(\rho) \qquad (1)$$

where $S$ is *Von Neumann entropy* [22] and $\rho_{diag}$ stands for the diagonal matrix extracted from the density matrix: ρ. Alternatively, the intuitive $l_1$-*norm of coherence* ($C_{l_1}$) is the sum of the absolute values of all off-diagonal elements of $\rho$:

$$C_{l_1}(\rho) = \sum_{\substack{i,j \\ i \neq j}} |\rho_{ij}|. \qquad (2)$$

Here $\rho_{ij}$ denotes an element from the i[th] row and j[th] column of $\rho$: $\rho_{ij} = \langle i|\rho|j\rangle$. Nonetheless, $C_{r.e}$ is the most illustrious contender as of now with strong operational meaning [2,23], quantification of coherence by $C_{r.e}$ is affected while $\rho$ is non-diagonalizable [24] or if the absolute value of any element of $\rho$ are close to "zero" (as $ln|\rho_{ij}| \to -\infty$ when $|\rho_{ij}| \to 0$). Moreover, the response of $C_{l_1}$ is not always adequate [25]. Hence, an accurate and upfront quantifier is in demand. In this regard, we present a function ($C_{PDD}$) that is superior to $C_{r.e}$ and $C_{l_1}$ in a pure state regime and easy to evaluate as only the diagonal elements of $\rho$ (i.e., $\rho_{ii}$: $\rho_{ii} = \langle i|\rho|i\rangle$) are in use.

The motivation behind achieving this function of { $\rho_{ii}$} is as follows: $l_1$-*norm of coherence* ($C_{l_1}$) is determined solely by the *absolute values* of the off-diagonal elements of $\rho$, implying that no phase term is involved. For example, let us consider a single qubit state with a state vector: $[\alpha * \exp(i\varphi_\alpha), \beta * \exp(i\varphi_\beta)]^T$ with $\alpha, \beta \in [0,1]$. The absolute value of the density-matrix off-diagonal element in the upper triangle: $|\alpha\beta * \exp(i(\varphi_\alpha - \varphi_\beta))| = \alpha\beta$, determines the amount of coherence ($\varphi_\alpha$, $\varphi_\beta$ not engaged). Interestingly, $\alpha^2$ and $\beta^2$ are principal-diagonal elements of the density-matrix, thus there must be a function (or functions) involving $\{\rho_{ii}\}$ that could be exploited as a coherence quantifier.

Most of the *distance-based* coherence quantifiers, like $C_{r.e}$ and $C_{l_1}$, measure the minimum distance of a quantum state from the set of *free states:* $\delta = \sum_{i=1}^{N} p_i |i\rangle\langle i|$, where $\{p_i\}$ are the mixing probabilities of the respective basis states $\{|i\rangle\}$ in an *N*-dimensional *Hilbert space*. Determining the free state closest to the measuring state is not always straightforward [3]. However, $C_{PDD}$ is devoid of such a problem as it uses *maximally coherent state*: $|\Psi_{max}\rangle = \frac{1}{\sqrt{N}}\sum_{i=1}^{N}|i\rangle$ as the reference point, instead of $\delta$ where all

$|\Psi_{max}\rangle$ of a particular dimension (N) have the same diagonal form in density matrix representation: $\rho_{diag}^{\Psi_{max}} = \frac{1}{N}\sum|i\rangle\langle i|$.

This letter is structured as follows: In the next section, we will define $C_{PDD}$ (*Theorem 1*), the distance of a pure state from the maximally coherent state (*Theorem 2*) and talk about the operating region of $C_{PDD}$ (*Theorem 3*). Subsequently, we discuss the properties that $C_{PDD}$ holds (especially prove monotonicity using the majorization theorem [26]), followed by validate $C_{PDD}$ being a proper coherence quantifier through comparing the evolution of $C_{PDD}$ with $C_{r.e}$ and $C_{l_1}$ (To do this, we take an isolated two-qubit system and perturb it with a laser pulse). Finally, we end our discussion with concluding remarks and future direction.

*Principal diagonal difference of coherence ($C_{PDD}(\rho)$).—*
**Theorem 1.** $C_{PDD}(\rho)$ *is defined as the sum of the absolute values of the difference between two principal diagonal elements (from the set: $\{\rho_{ii}\}$) of any pure-state density matrix ($\rho$) in all possible ways, normalized (to one) by dividing it by "$2*(N-1)$" and the whole subtracted from "1". $C_{PDD}(\rho)$ defines the superposition of a pure quantum state*

$$C_{PDD}(\rho) = 1 - \frac{\sum_{i,j}|\rho_{ii}-\rho_{jj}|}{2*(2^n-1)} = 1 - \frac{\sum_{i,j}|\rho_{ii}-\rho_{jj}|}{2*(N-1)}. \quad (3)$$

Here, $1 \leq i,j \leq N$; '$N$' ('$n$') is the dimension (number of qubits) of (in) the Hilbert space.

**Proof:** Upon inspection of the principal-diagonal elements of any normalized *pure-state* density operator, we come out with two important observations.

*Observation 1* for any maximally coherent state, all the elements of $\{\rho_{ii}\}$ are equal.

*Observation 2* for any incoherent state, only one element among the set of $\{\rho_{ii}\}$ is equal to 'one', and the rest of all are "zero."

Now, let us apply these observations to the absolute value of the difference between any two diagonal elements: $|\rho_{ii} - \rho_{jj}|$, taken from the set: $\{\rho_{ii}\}$. "*Observation 1*" achieves in the case of maximally coherent states that all such $\binom{N}{2}$ differences give "*zero*", that is,

$|\rho_{11} - \rho_{22}| = |\rho_{11} - \rho_{33}| = \cdots = |\rho_{11} - \rho_{NN}| = |\rho_{22} - \rho_{33}| = \cdots = |\rho_{(N-1)(N-1)} - \rho_{NN}| = 0 \quad (4.\text{A})$

Or, $\quad \sum_{i,j}|\rho_{ii} - \rho_{jj}| = 0 \quad (4.\text{B})$

Whereas, for an incoherent state, "*Observation 2*" implies that (N-1) members of the set: $\{|\rho_{ii} - \rho_{jj}|\}$ give "*one*." Let $\rho_{11} = 1$, so

$|\rho_{11} - \rho_{22}| = |\rho_{11} - \rho_{33}| = \cdots = |\rho_{11} - \rho_{NN}| = 1 \quad (4.\text{C})$

and the remaining '$\binom{N}{2} - (N-1)$' differences give "*zero*." Therefore, for an incoherent pure state

$$\sum_{i,j}|\rho_{ii} - \rho_{jj}| = (N-1). \quad (4.\text{D})$$

So far, we have considered all the combinations of any two different elements from the stack of $\{\rho_{ii}\}$. Thus, the total number of such combinations is $\binom{N}{2}$. However, because $\sum_{i,j}|\rho_{ii} - \rho_{jj}|$ contains both $|\rho_{ii} - \rho_{jj}|$ and $|\rho_{jj} - \rho_{ii}|$ (for a particular "i" and "j"), $\sum_{i,j}|\rho_{ii} - \rho_{jj}|$ is the sum of $2*\binom{N}{2}$ number of elements. As $|\rho_{ii} - \rho_{jj}| = |\rho_{jj} - \rho_{ii}|$, we should modify Eqn. (4. D) as follows:

$$\sum_{i,j}|\rho_{ii} - \rho_{jj}| = 2*(N-1). \quad (4.\text{E})$$

The summation: "$\sum_{i,j}|\rho_{ii} - \rho_{jj}|$" imparts an important operational meaning (see Theorem 2). It is bound by two extremes, however, in an opposite sense from the coherence quantifier point of view: $0 \leq \sum_{i,j}|\rho_{ii} - \rho_{jj}| \leq 2*(N-1)$. It becomes the *minimum* once the state is *maximally coherent* and achieves the *upper limit* for an *incoherent* state. Thus, it is not hard to admit that the normalized form of "$\sum_{i,j}|\rho_{ii} - \rho_{jj}|$" which is "$\frac{\sum_{i,j}|\rho_{ii}-\rho_{jj}|}{2*(N-1)}$" (normalized to *one*), subtracted from "*one*" acts as a self-normalized coherence quantifier, and that is the expression of $C_{PDD}(\rho)$ (Eqn. (3)).

The above discussion offers an important formula for the distance measure of a pure state from any maximally coherent state.

**Theorem 2** *For a fixed basis representation in an N-dimensional Hilbert space, the sum of the absolute values of the difference between any two principal diagonal elements of any pure-state density matrix ($\rho$) in all possible ways and divided by "$2*(N-1)$", measures the normalized distance of that state from any maximally coherent state ($\rho_{max}$)*

$$D(\rho, \rho_{max}) = \frac{\sum_{i,j}|\rho_{ii}-\rho_{jj}|}{2*(N-1)} \quad (5)$$

In the previous section, we already discussed the proof of this theorem. It is self-normalized to "one" ($0 \leq D(\rho, \rho_{max}) \leq 1$). $D(\rho, \rho_{max}) = 0$ for maximally coherent states (from Eqn. (4. B)) while $D(\rho, \rho_{max}) = 1$ for incoherent states (from Eqn. (4. E)), implying that as a quantum system approaches $\rho_{max}$, the magnitude of $D(\rho, \rho_{max})$ decreases. In the operational sense, '$D(\rho, \rho_{max})$' determines how much coherence of a pure state ($\rho$) is required to achieve the maximum coherence ($\rho_{max}$).

**Theorem 3** *The coherence quantifier, $C_{PDD}$, cannot be applied directly to any mixed-state density operator.*

**Proof** Consider a mixed state in an N-dimensional Hilbert space that is a mixture of two pure states $|\Psi_1\rangle = \sum_{i=1}^{N}\Psi_1^i|i\rangle$ and $|\Psi_2\rangle = \sum_{i=1}^{N}\Psi_2^i|i\rangle$ with mixing probability $p_1$ and $p_2$, respectively. $\{|i\rangle\}$ is the set of computational basis states. $\Psi_1^i = |\Psi_1^i|e^{k\varphi_{\Psi_1^i}}$ and $\Psi_2^i = |\Psi_2^i|e^{k\varphi_{\Psi_2^i}}$ are the complex coefficients ($k = \sqrt{-1}$, $\varphi_{\Psi_{1,2}^{i,j}}$ are the angles of coefficients in the complex plane); The mixed state in density matrix form is as follows:

$$\rho_{\Psi_1,\Psi_2} = \sum_{i,j=1}^{N}[\,p_1\Psi_1^i\Psi_1^{j*}|i\rangle\langle j| + p_2\Psi_2^i\Psi_2^{j*}|i\rangle\langle j|\,] \quad (6.\text{A})$$

The element of $i^{th}$ row and $j^{th}$ column of $\rho_{\Psi_1,\Psi_2}$ is

$$\rho_{\Psi_1,\Psi_2}^{(i,j)} = p_1\Psi_1^i\Psi_1^{j*} + p_2\Psi_2^i\Psi_2^{j*} \quad (6.\text{B})$$

Use of the expanded forms of the state coefficients $\Psi_1^i$ and $\Psi_2^i$ on the above equation gives:

$$\rho_{\Psi_1,\Psi_2}^{(i,j)} = p_1|\Psi_1^i||\Psi_1^j|e^{k(\varphi_{\Psi_1^i}-\varphi_{\Psi_1^j})} + p_2|\Psi_2^i||\Psi_2^j|e^{k(\varphi_{\Psi_2^i}-\varphi_{\Psi_2^j})} \quad (6.\text{C})$$

From Eqn. (6. C), we get the absolute value of $\rho_{\Psi_1,\Psi_2}^{(i,j)}$ with the help of complex algebra, which becomes

$$|\rho_{\Psi_1,\Psi_2}^{(i,j)}| = \sqrt{\begin{array}{c} p_1^2|\Psi_1^i|^2|\Psi_1^j|^2 + p_2^2|\Psi_2^i|^2|\Psi_2^j|^2 \\ +2p_1p_2\sqrt{|\Psi_1^i|^2|\Psi_1^j|^2|\Psi_2^i|^2|\Psi_2^j|^2} \\ f\left(\varphi_{\Psi_{1,2}^{i,j}}\right) \end{array}} \quad (6.\text{D})$$

Here,

$$f\left(\varphi_{\Psi_{1,2}^{i,j}}\right) = \cos\left(\varphi_{\Psi_1^i} - \varphi_{\Psi_1^j}\right)\cos\left(\varphi_{\Psi_2^i} - \varphi_{\Psi_2^j}\right) + \sin(\varphi_{\Psi_1^i} - \varphi_{\Psi_1^j})\sin(\varphi_{\Psi_2^i} - \varphi_{\Psi_2^j}) \quad (6.\text{E})$$

For a particular set of diagonal elements (here $\{p_1|\Psi_1^i| + p_2|\Psi_2^i|\}$, with the integer "i" going from 1 to N) in any order, the sum of off-diagonal elements of a density matrix should be unique to qualify a function of the set of principal diagonal elements as a coherence quantifier. The same set of principal diagonal elements, however, assigns multiple values for a $|\rho_{\Psi_1,\Psi_2}^{(i,j)}|$ (with $i \neq j$) in a mixed state. These multiple possibilities are due to the presence of the term: $f\left(\varphi_{\Psi_{1,2}^{i,j}}\right)$ in the expression of $|\rho_{\Psi_1,\Psi_2}^{(i,j)}|$ (see Eqn. (6. D)). As off-diagonal elements are solely responsible for quantum coherence, different coherence amounts attributed to the same coherence measure are unacceptable. This is the reason we cannot apply $C_{PDD}$ directly to mixed states.

For any pure state (let say $p_2 = 0$ and $p_1 = 1$), Eqn. (6. D) will be as follows:

$$|\rho_{\Psi_1}^{(i,j)}| = p_1^2|\Psi_1^i|^2|\Psi_1^j|^2. \quad (6.\text{F})$$

Interestingly, Eqn. (6. F) shows that the absolute value of any off-diagonal element of a pure state is not associated with any phase term, thus $C_{PDD}$ is unique to any principal diagonal set $\{|\Psi_1^i|^2\}$ and for that reason $C_{PDD}$ can measure coherence correctly for a *pure state* but provides incorrect output to a *mixed state*. However, the correct quantification for this case is feasible if we know the involved *pure states* and their *mixing probabilities* apriori.

*Properties of $C_{PDD}$.—*
**Non-negativity.** Since $0 \leq D(\rho,\rho_{max}) \leq 1$, it is easy to show that $0 \leq C_{PDD} \leq 1$ too. While the pure state is one of the basis states: $\{|i\rangle\}$, it is *incoherent*. We have previously seen that for the basis state, $D(\rho,\rho_{max}) = 1$, which makes $C_{PDD} = 0$, as $C_{PDD} = 1 - D(\rho,\rho_{max})$; Again, from this, we can infer that a minute deviation from the basis state makes $C_{PDD} > 0$. Therefore, the first criterion of being a valid coherent measure [1–3] in a stronger sense, i.e., $C_{PDD}(\rho) = 0$ iff $\rho \in \delta$, is satisfied.
**Self-normalization.** By definition $C_{PDD}$ is normalized to 'one' and $C_{PDD}(\rho) = 1$ only if all the diagonal elements are the same (*maximally coherent state*); otherwise, $C_{PDD}(\rho) < 1$. Whereas other quantifiers, like the $l_1$-norm of coherence and the relative entropy of coherence need to normalize [27], the $C_{PDD}$ is self-normalized by its definition.
**Monotonicity.** Monotonicity under incoherent operation is a necessary criterion [1,2,4] for any valid coherence quantifier. With the help of the majorization theorem [26], Shuanping Du et al. were able to show that the Winger-Yanase-Dyson skew information [20] does not obey the monotonicity criterion. Here we use the same approach to prove that $C_{PDD}$ is a monotonic function.

*Proof* Let us take two arbitrary pure states: $|\Psi_1\rangle$ and $|\Psi_2\rangle$ in an N-dimensional Hilbert space, whose $\{|i\rangle\}$ basis representation is the same as used in the proof of "Theorem 3." Now, the majorization theorem says that $|\Psi_1\rangle$ is transformable to $|\Psi_2\rangle$ through incoherent operation if and only if $|\Psi_2\rangle$ majorizes $|\Psi_1\rangle$, i.e.,

$$\sum_{i=1}^{l} P^\downarrow(\Psi_2^i) \geq \sum_{i=1}^{l} P^\downarrow(\Psi_1^i) \quad (7.\text{A})$$

for all $l \in \{1,\ldots,N\}$ where $P(\Psi_m^i) = |\Psi_m^i|^2$ (with $m \in \{1,2\}$) and '$\downarrow$' indicates the decreasing order of the diagonal elements in $\{P(\Psi_m^i)\}_{i=1}^N$. Because incoherent operations can never increase coherence ($C(\rho) \geq C(\Lambda(\rho))$; $\Lambda$: arbitrary incoherent operation in the Hilbert space of $\rho$), $C_{PDD}(|\Psi_1\rangle)$ must be greater than or equal to $C_{PDD}(|\Psi_2\rangle)$ for $|\Psi_1\rangle \xrightarrow{\Lambda} |\Psi_2\rangle$ in order to be possible.

Case 1: N=2
Let $\{P(\Psi_1^i)\} = \{a, 1-a\}^T$ and $\{P(\Psi_2^i)\} = \{b, 1-b\}^T$ with $a, b > 1/2$ (maintaining decreasing order). The superscript, "T", represents the transpose of a row vector. If $|\Psi_1\rangle \xrightarrow{\Lambda} |\Psi_2\rangle$, then $b \geq a$ (majorization principle). From Eqn. (3) we have

$$C_{PDD}(|\Psi_1\rangle) = 1 - |2a-1|$$
$$\text{and} \quad C_{PDD}(|\Psi_2\rangle) = 1 - |2b-1|. \quad (7.\text{B})$$

So, by applying majorization inequality (i.e., $b \geq a$) to the above equation (Eqn. (7. B)), it is obvious that $C_{PDD}(|\Psi_1\rangle) \geq C_{PDD}(|\Psi_2\rangle)$.

Case 2: N=3
Suppose, $\{P(\Psi_1^i)\} = \{a_1, a_2, 1-a_1-a_2\}^T$ and $\{P(\Psi_2^i)\} = \{b_1, b_2, 1-b_1-b_2\}^T$ in decreasing order and again $|\Psi_1\rangle \xrightarrow{\Lambda} |\Psi_2\rangle$ which implies $b_1 \geq a_1$ and $b_1 + b_2 \geq a_1 + a_2$. From Eqn-3, we get the coherence amounts as follows:

$$C_{PDD}(|\Psi_1\rangle) = 1 - \frac{|a_1-a_2|+|2a_1+a_2-1|+|a_1+2a_2-1|}{2}$$
$$\text{And } C_{PDD}(|\Psi_2\rangle) = 1 - \frac{|b_1-b_2|+|2b_1+b_2-1|+|b_1+2b_2-1|}{2}. \quad (7.\text{C})$$

Since $a_1 > a_2$ and $b_1 > b_2$, all subtractions moduli have become normal subtractions, i.e.,

$$\frac{|a_1-a_2|+|2a_1+a_2-1|+|a_1+2a_2-1|}{2} = a_1 + (a_1+a_2) - 1$$
$$\text{And } \frac{|b_1-b_2|+|2b_1+b_2-1|+|b_1+2b_2-1|}{2} = b_1 + (b_1+b_2) - 1. \quad (7.\text{D})$$

By using the majorization inequalities (i.e., $b_1 \geq a_1$ and $b_1 + b_2 \geq a_1 + a_2$) on Eqn. 7.D and its outcome on Eqn. (7. C), we can easily find that $C_{PDD}(|\Psi_1\rangle) \geq C_{PDD}(|\Psi_2\rangle)$.

According to the above results, for two- and three-dimensional cases, $C_{PDD}$ obeys monotonicity. Therefore, by the induction method, it is obvious that $C_{PDD}$ is monotonic for any finite value of "N." (proved)

N.B.: *strict monotonicity* and *convexity* are other coherence criteria but have no direct proof for $C_{PDD}$ (as $C_{PDD}$ acts correctly only for the pure-state regime). Nonetheless, we will provide an indirect account of proof by comparing the evolution pattern of $C_{PDD}$ to $C_{r.e}$ and $C_{l_1}$.

*Validation of $C_{PDD}$ (the numerical analysis).*—In previous sections, we explained how the principal diagonal difference of coherence ($C_{PDD}$) is bound by *incoherent* and *maximally coherent* states and proved its monotonicity under incoherent operation. To put this theory on solid ground and determine whether the response of $C_{PDD}$ is better than $C_{l_1}$ and $C_{r.e}$ or not, we use the numerical evolution of $C_{PDD}$ along with $C_{l_1}$ and $C_{r.e}$.

*Methodology.* Coherence changes due to the variation in basis-state superposition. In other words, state coherence evolves with basis-state populations, accordingly. Here, we go by a simple but well-accepted, semi-classical approach to *laser pulse-qubit (qubits) interaction* [28,29] (to oscillate state populations), where the laser pulse (considered classical, without loss of generality [30–32]) makes the *Hamiltonian* of the qubit-system (quantum) time (or pulse area) dependent. In this case, we think of an isolated two-qubit system (no decay loss) and simulate the evolution of the qubit system upon interaction with the laser pulse. To ease calculation, we use the *FM-transformed* form [29,33] of the *perturbed Hamiltonian*:

$$H_i = \begin{pmatrix} \Delta_i & \Omega_i/2 \\ \Omega_i^*/2 & 0 \end{pmatrix} \quad (8)$$

where $i$ takes the value 1 (or 2) for *qubit-1* (or *qubit-2*); $\Delta_i$ and $\Omega_i$ are *detuning* and *Rabi frequency* [28,29,34] of the i$^{th}$ qubit, respectively. Here, we study two cases: (a) *qubit-qubit non-interacting:* $H = H_1 + H_2$; and (b) *qubit-qubit interacting:* $H = H_2 \otimes H_1$; keeping interaction strength "one." For each one of these cases, we examine both *resonance ($\Delta_i = 0$)* and *detuning ($\Delta_i \neq 0$)* conditions. '$H$' is the total *Hamiltonian* of the *two-qubit system*. The governing equation used here is the *Liouville equation* [30,35]: $\dot{\rho} = i\hbar^{-1}[\rho, H]$. To compare coherence evolutions, we use normalized forms of $C_{r.e}$ and $C_{l_1}$ [27].

*Results and Discussions.* Figures 1 and 2 depict the evolution of $C_{PDD}$ relative to $C_{l_1}$ and $C_{r.e}$. As both $C_{l_1}$ and $C_{r.e}$ fulfill all four necessary and sufficient coherence criteria, the similar evolution patterns of $C_{PDD}$ for each of the *non-interacting* (FIGS. 1(a) and 2(a)) and *interacting* (FIGS. 1(b) and 2(b)) cases substantiate that it is a good coherence quantifier. The *resonant and non-interacting* plot (FIG. 1(a)) shows that while all the population curves meet at the same point (i.e., *maximally coherent states*), all three coherence quantifies reach "1"; likewise, wherever only the state $|11\rangle$ achieves the population "one" and all remaining state populations stay at "0" (implies *incoherent state*), all the quantifiers become "0," and in between these two extremes, they follow the condition $0 < C(\rho) < 1$; that supports two properties for $C_{PDD}$: (a) $0 \leq C_{PDD}(\rho) \leq 1$; (b) $C_{PDD}(\rho) = 0$ (or 1) only when the state is *incoherent* (or *maximally coherent*). FIG. 1.b displays the *resonance* case with *qubit-qubit interaction (entangled)*, keeping the interaction strength at unity. The frequency of population oscillation and the corresponding coherence oscillation, in this case, increases, and the *maximally coherent state* is never achieved, as expected.

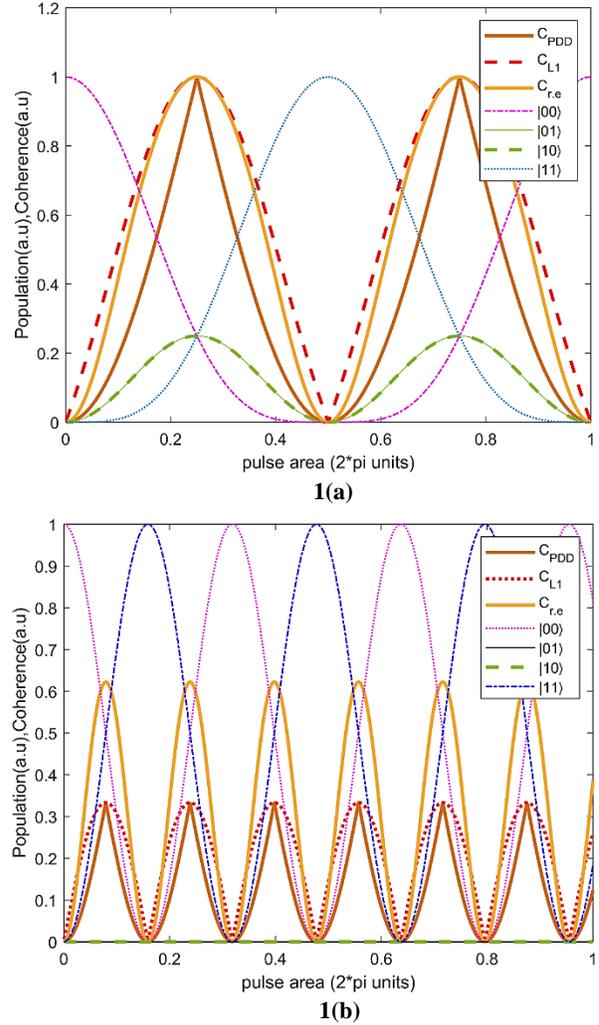

FIG. 1: A comparison of the evolutions (Vs. laser-pulse area) of coherence quantifiers ($C_{PDD}$, $C_{r.e}$, and $C_{l_1}$) for an isolated two-qubit system perturbed by a laser pulse with *resonant* pulse-qubit interactions ($\Delta_i = 0$; $\Delta_i$: detuning of the i$^{th}$-qubit for $i = \{1,2\}$); The evolutions of the basis-state populations (indicated by-$|00\rangle$, $|01\rangle$, $|10\rangle$ and $|11\rangle$) are also linked in order to determine the maximum (or minimum) superposition points. **Figs. 1(a)** and **1(b)** represent qubit-qubit *non-interacting* and *interacting* cases, respectively.

FIG. 2 exhibits the *detuning* case, and we apply different detuning for *qubit-1* and *qubit-2* to create some amount of asymmetry within the result. To determine which of the three quantifiers is best, we look for the situations in which the *populations* meet at a point (the *maximally coherent state*) or very close to it (highlighted by a circular boundary connected to the arrow tail) and verify the corresponding responses of these three quantifiers (highlighted by a circular boundary connected to the arrowhead). Observations show that $C_{l_1}$ is *least sensitive* to a small change in superposition (coherence) in the region surrounding maximally coherent states, whereas

$C_{PDD}$ is the *most sensitive* and $C_{r.e}$ is in between but closely follows $C_{l_1}$. For example, in FIGS. 2(a) and 2(b), *arrow 1* points to the response when the state is maximally coherent with negligible deviation, and *arrow 2* indicates the response when the state deviates more than in the first case. The responses show that the change in $C_{l_1}$ and $C_{r.e}$ is negligible for the visible change in state superposition, whereas the changes in $C_{PDD}$ occur in a *linear* (according to our perception) and *sensible* manner.

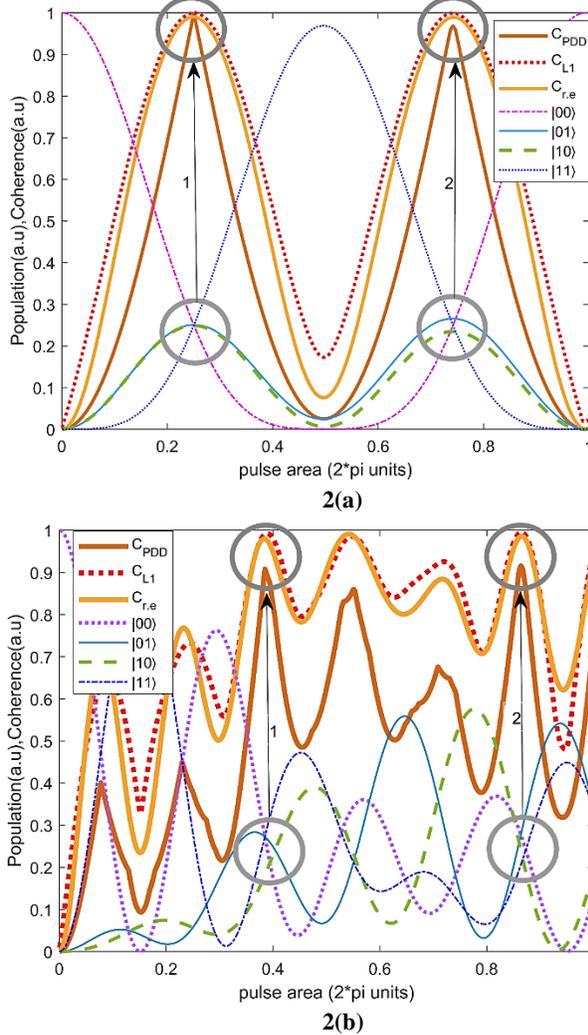

FIG. 2: A comparison of the evolutions (Vs. laser-pulse area) of coherence quantifiers ($C_{PDD}$, $C_{r.e}$, and $C_{l_1}$) for an isolated two-qubit system perturbed by a laser pulse with *off-resonant* pulse-qubit interactions ($\Delta_1 = 0.5\Omega_R, \Delta_2 = \Omega_R$; $\Delta_i$: i$^{th}$-qubit detuning ($i = \{1,2\}$); $\Omega_R$: Rabi frequency); The evolutions of basis-state populations (indicated by- $|00\rangle, |01\rangle, |10\rangle$ and $|11\rangle$) are also linked in order to determine the maximum (or minimum) superposition zones. **Figs. 2(a)** and **2(b)** represent qubit-qubit *non-interacting* and *interacting* cases, respectively. Each of the plots shows two arrows (denoted by "1" and "2"); circles connecting to the arrowheads highlight coherence responses corresponding to the circles highlighting "close to maximal coherence" situations connecting to the arrow-tails.

From all these plots, it is quite clear that the response of both $C_{r.e}$ and $C_{l_1}$ is poor in the neighborhood region of *maximally coherent states*, whereas $C_{PDD}$ shows a *linear* response to the change in state superposition (coherence), which strengthens the statement that $C_{PDD}$ *is* the exact measure of superposition (*Theorem 1*).

*Remarks.—* (i) In this letter, we introduced a new expression for the coherence (or superposition) quantifier: $C_{PDD}$, that depends only on the diagonal elements of the density matrix represented in the measurement basis; demonstrated its *positivity, self-normalizability,* and *monotonicity* under any incoherent operation; and solidified this theory through the numerical evolution of $C_{PDD}$ in comparison with $C_{r.e}$ and $C_{l_1}$. (ii) Given the set of diagonal elements, whereas the majorization theorem [26] can only compare two pure states in terms of coherence, $C_{PDD}$ directly provides the total amount of coherence which signifies that the set of principal diagonal elements of a pure state density-matrix possesses the complete information regarding its coherence (or superposition) measure. (iii) The linear change in $C_{PDD}$ with the change of *superposition* imparts that $C_{PDD}$ is a proper definition of *quantum superposition*.

*Future direction.—* $C_{PDD}$ can precisely measure quantum superposition (coherence), but we cannot apply it directly to a mixed state as it only involves the diagonal elements of a density matrix. To solve this problem, we can find a mathematical relationship between $C_{PDD}$ and any other quantifier (that works in both pure and mixed-state regimes) that allows us to create another function that works as well as $C_{PDD}$, and additionally, covers both pure and mixed state regimes, like other coherence quantifiers.

*Acknowledgment:*
We thank S. Goswami for language correction and editing. DG acknowledges funding support from MEITY, SERB, and STC ISRO of the Govt. of India.